\documentclass[pra,twocolumn,aps,superscriptaddress,longbibliography]{revtex4-1}

\usepackage{bm}%
\usepackage{float}
\usepackage{lipsum} 
\expandafter\ifx\csname package@font\endcsname\relax\else
 \expandafter\expandafter
 \expandafter\usepackage
 \expandafter\expandafter
 \expandafter{\csname package@font\endcsname}%
\fi

\usepackage{array}
\usepackage{amsmath,amssymb}
\usepackage{MnSymbol}
\usepackage{tikz-feynman,contour} 
\usetikzlibrary{shapes,arrows,positioning,automata,backgrounds,calc,er,patterns}
\usepackage{feynmf}
\tikzfeynmanset{compat=1.0.0} 

\usepackage{graphicx}
\usepackage{mathrsfs}
\usepackage{bm}
\usepackage{mathtools}
\usepackage{epstopdf}
\usepackage[breaklinks]{hyperref}

\hypersetup{%
   pdfpagemode=None, 
   pdfstartpage=1,
   pdfmenubar=true,
   pdftoolbar=true,
   colorlinks = true,
   linkcolor=blue,
   citecolor=blue,
   urlcolor=blue,
   bookmarksopen=false
 }
 \usepackage{amsmath}
\usepackage{amsfonts}
\usepackage{mathtools}
\usepackage{graphicx}
\usepackage{fixme}
\usepackage{color}
\usepackage{cancel}

\DeclareUnicodeCharacter{2212}{-}

\begin{document}

\title{A mobile ion in a Fermi sea}




\author{Esben Rohan Christensen}
\affiliation{Center for Complex Quantum Systems, Department of Physics and Astronomy, Aarhus University, Ny Munkegade 120, DK-8000 Aarhus C, Denmark.}
\author{Arturo Camacho-Guardian}
\affiliation{T.C.M. Group, Cavendish Laboratory, University of Cambridge, JJ Thomson Avenue, Cambridge, CB3 0HE, U.K.}
\author{Georg M. Bruun} 
\affiliation{Center for Complex Quantum Systems, Department of Physics and Astronomy, Aarhus University, Ny Munkegade 120, DK-8000 Aarhus C, Denmark.}
\affiliation{Shenzhen Institute for Quantum Science and Engineering and Department of Physics, Southern University of Science and Technology, Shenzhen 518055, China. }

\begin{abstract}
The remarkable single particle control of individual ions combined with the versatility of ultracold atomic gases makes hybrid ion-atom system an exciting new platform for 
quantum simulation of few- and many-body quantum physics. Here, we study theoretically the  properties of a mobile ion immersed  in a quantum degenerate  gas of fermionic atoms.
Using an effective low-energy atom-ion interaction together with a well established approach that includes exactly two-body correlations, we calculate the full 
spectral response of the ion and demonstrate the existence of several quasiparticle branches, which are  charged analogues of the Fermi polaron observed in neutral atomic gases. Due to the long-range 
nature of the atom-ion interaction, these ionic Fermi polarons have several  properties distinct from their neutral counterparts such as the simultaneous presence of several stable states and 
smooth transitions from repulsive to attractive polarons with increasing interaction strength. Surprisingly, the residue of the ionic polaron is  shown to increase 
with the Fermi density for fixed interaction strength, which is in marked contrast to the neutral polaron. The properties of the ionic polaron approach that of the neutral 
polaron only in the low density limit where the average interparticle spacing
is larger than the characteristic length of the atom-ion interaction. We finally analyse the effects of the Fermi gas on the molecular ions, 
which are bound atom-dimer states. 
 \end{abstract}

\date{\today}
\maketitle

\section{Introduction}
The ability to prepare and  manipulate both single ions as well as neutral atomic gases has opened up  opportunities to explore quantum  systems consisting of 
ions embedded in a neutral gas environment~\cite{Tomza2019}. 
Due to their hybrid ion-neutral nature, such systems represent a new powerful quantum simulation platform for  investigating fundamental physics ranging from two-body collisional 
processes, molecule formation, to complex many-body states, and they may also have applications for future quantum information technologies. 
While the field of hybrid ultracold quantum ion-atom systems is still in its infancy, experiments have already explored 
ion-atom collisions and sympathetic cooling~\cite{Grier2009,Zipkes2010,Feldker2020,Schmidt2020}, three-body recombination, molecule formation, charge transfer 
reactions~\cite{Harter2012,Ratschbacher2012,Sikorsky2018,Dieterle2020b,Mohammadi2021}, ions in a Bose-Einstein condensate (BEC)~\cite{Kleinbach2018}, charge transport~\cite{Dieterle2021}, 
and the use of ions for high resolution microscopy~\cite{Veit2021}. Theoretically, one has used a wide range of techniques to study the 
few-body properties such as atom-ion scattering, three-body recombination and  molecule formation, cold chemistry, as well as  sympathetic cooling~\cite{Lukin2002,Gao2010,Krukow2016,Schurer2017,Perez2021,oghittu2021}. 
The many-body properties of these hybrid systems in the quantum degenerate regime 
have also been investigated theoretically with an important early exploring a  static ion in a Bose-Einstein condensate (BEC) using the Gross-Pitaevskii equation~\cite{Massignan2005}. Later,
 calculations based on a weak coupling Fr\"ohlich model~\cite{Casteels2011}, Monte-Carlo numerics~\cite{Astrakharchik:2021wl} as well as a variational approach~\cite{Christensen2021}, 
predicted that immersing a mobile ion in a BEC can lead to the formation of  quasiparticles, which are the charged analogue of the Bose polaron observed for neutral impurity atoms in a 
BEC~\cite{Jorgensen2016,Hu2016,Ardila2019,Yan2020}. The atom-ion interaction is however much longer ranged than that between neutral atoms, and it is therefore 
presently unclear under which conditions this leads to strong three-body losses that may complicate the observation of such ionic Bose polarons~\cite{Harter2012,Mohammadi2021}. 

 Immersing an ion in a single-component Fermi gas generally provides a more stable system since the Pauli exclusion principle suppresses three-body losses 
 involving two identical fermions. As such,  this 
provides a promising pathway to realise  interesting quantum states such as  charged quasiparticles in analogy with the observation of  stable Fermi polarons formed by neutral impurity atoms in Fermi gases~\cite{Schirotzek2009,Kohstall12,Koschorreck12,Scazza2017,Fritsche2021}, where the interaction between the impurity and the environment is short-range~\cite{pethick}.
Fermi polarons have recently also been observed in the context of exciton-polaritons~\cite{Sidler2016}, which are again well described theoretically using a short-range 
interaction~\cite{Efimkin2017,Tan2019,Bastarrachea2021}. The properties of  Fermi polarons with long-range interactions remain on the other hand much less explored.

Motivated by this,  we  explore here the properties of a single mobile ion in a quantum degenerate  gas of fermionic atoms. Using a low 
energy  effective interaction combined with the simplest possible approach that includes exactly 
two-body atom-ion correlations known to give quantitative accurate results for neutral impurities in a Fermi gas, we calculate the full spectral response of the ion. 
We show that there are several quasiparticle branches corresponding to ionic Fermi polarons, which due to the long-range atom-ion interaction have  properties quite different from the neutral Fermi polaron. 
This includes  the simultaneous presence of several stable polaron branches and the smooth transition from repulsive to attractive polarons with increasing interaction strength.  Remarkably,   
 the residue of the attractive polaron is demonstrated to increase with increasing Fermi density and remains non-zero in the so-called BEC limit in stark contrast to the residue of the neutral attractive polaron. 
Only when the density is so low that the average distance between the particles is larger than the characteristic length of the atom-ion interaction do the properties of the ionic polaron 
resemble those of the neutral polaron. We finally analyse the effects of the Fermi gas on the bound atom-ion states corresponding to molecular ions.  

The manuscript is organised as follows. In Sec.~\ref{System}, we describe the hybrid atom-ion system, and in Sec.~\ref{twobody} we analyse the two-body atom-ion scattering and 
molecule formation using both a real space and a momentum space approach.  The many-body problem is then explored in Sec.~\ref{IonicPolaron} where  
the full spectral response of an ion in a Fermi gas is  calculated. In Sec.~\ref{molecule}, we discuss the formation of molecular ions in the presence of the Fermi sea and we conclude and provide an outlook in 
Sec.~\ref{Conclusion}.

  \begin{figure}[!htb]
\centering
    \includegraphics[width=\columnwidth,height=.8\columnwidth]{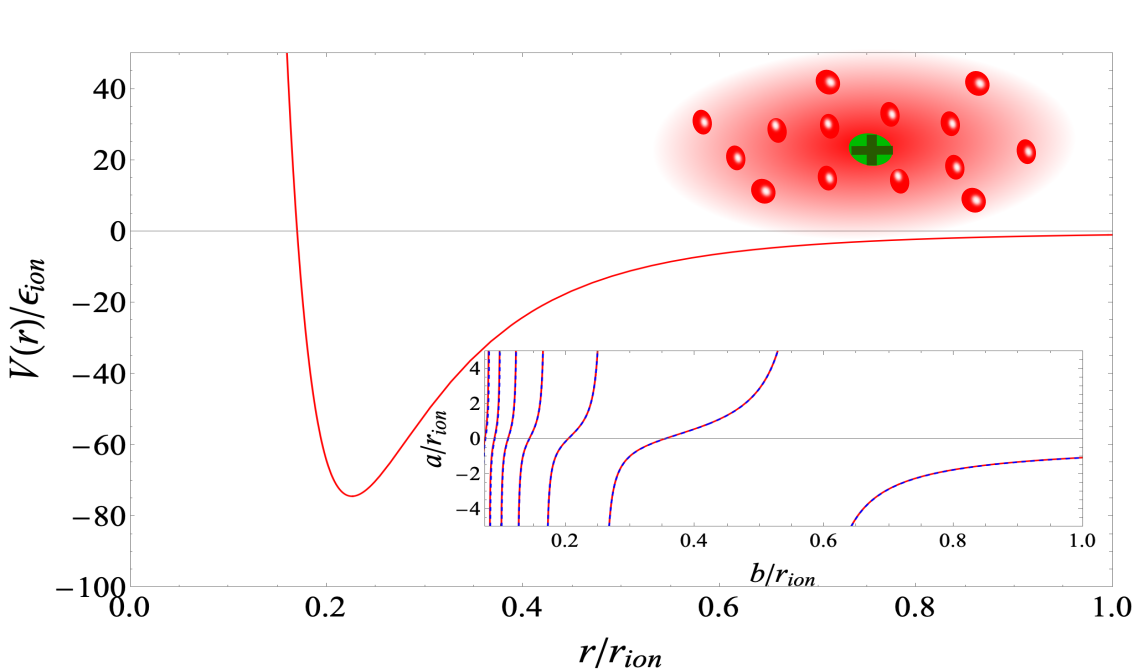}
    \caption{The ion-atom interaction potential as a function of $r$.   (Inset top) Illustration  of a single ion in  a single component Fermi gas. (Inset bottom) The atom-ion 
     scattering length  $a$ as a function of $b$ calculated by direct solution of the Schr\"odinger equation in real space (red line) and from the Lipmann-Schwinger equation (blue dashed line).}
    \label{scatlen2}
\end{figure}

\section{System}\label{System}
We consider a single mobile ion of mass $m$ immersed in a gas of single-component fermionic atoms of mass $m_F$ at zero temperature. The Hamiltonian  describing the system is 
  \begin{gather}
  \label{H1}
\hat{H} = \sum_{\mathbf{k}} \left(\varepsilon_{\mathbf k}\hat{a}^\dagger_\mathbf{k} \hat{a}_\mathbf{k} +\varepsilon_{F\mathbf k} \hat{c}^\dagger_\mathbf{k} \hat{c}_\mathbf{k}\right) + 
\frac1{\mathcal V}\sum_{\mathbf{k}, \mathbf{k'},\mathbf{q}} \!V(\mathbf{q}) \hat{c}^\dagger_{\mathbf{k'-q}} \hat{c}_{\mathbf{k'}} \hat{a}^\dagger_{\mathbf{k}+\mathbf{q}}
\hat{a}_\mathbf{k},
\end{gather}
where $\hat{a}^\dagger_\mathbf{k} $ and $\hat{c}^\dagger_\mathbf{k}$ create an ion and a fermion with momentum $\mathbf k$ and 
energy  $\varepsilon_{\mathbf k} =k^2/2m$ and $\varepsilon_{F\mathbf k}=k^2/2m_F$ respectively. 
The second term in Eq.~\eqref{H1} describes the ion-atom interaction and we can as usual  neglect the short-range interaction between the 
spin-polarised fermions.  Finally, $\mathcal V$ denotes the volume of the system and we take $\hbar=1$ throughout

\section{Two-Body properties}
\label{twobody}
Before exploring the many-body problem, we start by discussing the relevant low energy atom-ion two-body physics. For a large separation $r$, the dominant source of the interaction comes from the 
electric field of the ion polarising the atom. This  gives rise to an attractive interaction of the form $\sim-\alpha/r^4$
where $\alpha$ is proportional to the static electric polarisability of the atom and the charge of the ion. For short distances,  the overlap between electron clouds of the atom and the ion 
gives rise to strong short-range repulsion. The full atom-ion interaction is quite complicated with many deeply bound states~\cite{Tomza2019}.
Since we focus on the low energy physics, it is sufficient to use an effective 
interaction  of the form~\cite{Krych2015}
\begin{equation}
    V(r) = -\frac{\alpha}{(r^2+b^2)^2}\frac{r^2-c^2}{r^2+c^2}.
    \label{potential}
\end{equation}
Here, $b$ determines the depth of the potential, whereas $c$ determines distance below which the interaction is repulsive, see Fig.~\ref{scatlen2}. 
 The  strong repulsion at short distances is accounted for by the fact that  $V(0)=\alpha/b^4$ is much larger than any other relevant energy in the system.  
From the long-range behaviour $\sim-\alpha/r^4$ of the interaction, we can define the characteristic length $r_\text{ion}=\sqrt{2m_r\alpha}$ and energy $\varepsilon_\text{ion}=1/2m_rr_\text{ion}^2$
where $m_r=(m_Fm)/(m+m_F)$ is the reduced mass~\cite{Massignan2005}.

The parameters $b$ and $c$ can be determined by requiring that the effective interaction Eq.~\eqref{potential} reproduces the correct low energy 
two-body physics  relevant for the many-body problem at hand, i.e.\ 
the  scattering length and  energy of the highest  bound state of a given atom-ion combination~\cite{Krych2015,oghittu2021}. An ion with momentum $\mathbf k_1$ scattering on an atom with momentum $\mathbf k_2$  leading to the momentum transfer $\mathbf q$ 
 is described by the scattering matrix, which obeys Lippmann-Schwinger equation 
\begin{gather}
\mathcal{T}({\mathbf k}_1,{\mathbf  k}_2,{\mathbf q};E)=V({\mathbf q})+\frac1{\mathcal V}\sum_{\mathbf q'}V({\mathbf q}')\times\nonumber \\
 \mathcal G_2(\mathbf{ k}_1+\mathbf q',\mathbf{ k}_2-\mathbf q',E)\mathcal{T}(\mathbf{ k}_1+\mathbf{q}',{\mathbf  k}_2-\mathbf{q}',\mathbf{q-q}';E).
\label{LippmannSchwinger}
\end{gather}
Here, $E$ is the total energy and  
$\mathcal G_2^{-1}(\mathbf{ k}_1,\mathbf{ k}_2,E)=E-\varepsilon_{\mathbf k_1}-\varepsilon_{F\mathbf k_2}$ describes the propagation of 
an ion and an atom with momenta  $\mathbf k_1$ and $\mathbf k_2$ and total energy $E$. 
The $s$-wave scattering length can be obtained  by expanding the zero center of mass scattering matrix in Legendre polynomials 
$\mathcal T(\mathbf k, -\mathbf k,\mathbf q;E)=\sum_{l=0}(l+1/2)\mathcal T_{l}(k,q;E)P_{l}(\cos\theta_{\mathbf k,\mathbf k+\mathbf q})$ giving 
\begin{gather}
a=\frac{2\pi}{m_r}\mathcal T_0(0,0;0).
\end{gather}
We plot in Fig.~\ref{scatlen2} the  scattering lengths obtained from this procedure as a function of $b$.
For alkali atoms the interaction is repulsive for $r\lesssim 10 a_0 $ while $r_{\text{ion}} \sim \mathcal{O}(10^2)$nm and we therefore take $c=0.0023 r_\text{ion}$ in this manuscript. We see that the scattering length exhibits several divergencies as the atom-ion interaction potential deepens with decreasing $b$,
which are caused by the  emergence of $s$-wave bound dimer states. The energies $\varepsilon_M$ of these 
molecular ions can be found as poles of the scattering matrix as a function of the energy.


We also show in Fig.~\ref{scatlen2} the scattering obtained from solving the 
 Schr\"odinger equation $H\psi(r)=E\psi(r)$ in real space. 
 For vanishing energy and zero angular momentum, the radial wave function takes the form $\psi(r)=\chi(r)/r$, where $\chi(r) = c_1 r+ c_2$ for large $r$ and 
 from this one can extract the scattering length as $a = - c_2/c_1$. The excellent agreement between the scattering length obtained from this real space method and 
 from the Lippmann-Schwinger equation confirms the accuracy of our numerical calculations. We note that they are significantly more complicated than for the usual case of a short-range 
 interaction, since one needs to retain the full momentum dependence of the atom-ion interaction.

 \section{Ionic Fermi Polaron}
\label{IonicPolaron}
Now we turn our attention to the many-body problem of a mobile ion interacting with a single-component Fermi sea. The properties of an ion with momentum $\mathbf p$
 are described by the  Green's function 
\begin{align}
\mathcal G^{-1}(\mathbf p,i\omega)=i\omega-\varepsilon_{\mathbf p}-\Sigma(\mathbf p,i\omega),
\label{Gion}
\end{align}
where $\Sigma(\mathbf p,\omega)$ is the ion self-energy and $i\omega$ a Matsubara frequency. Inspired by its successful use for analysing neutral 
atomic Fermi gases, we use ladder approximation to calculate 
the  self-energy. This is  the simplest approach that includes  the  two-body atom-ion physics exactly in a many-body setting, and it gives  
\begin{align}
\Sigma(\mathbf p,i\omega)=\sum_{\mathbf q,\omega_q}\mathcal T(\mathbf p,\mathbf  q,{\mathbf 0};i\omega+i\omega_q)G_F(\mathbf q,i\omega_q), 
\label{Selfenergy}
\end{align}
where $G_F^{-1}(\mathbf q,i\omega_q)=i\omega_q-\varepsilon_{F\mathbf q}+\mu_F$ is the non-interacting Green's function of the atoms with $\mu_F$ their 
chemical potential. The scattering matrix  $\mathcal T$ in Eq.~\eqref{Selfenergy} is  obtained from the $s$-wave component of Eq.~\eqref{LippmannSchwinger} using 
$\mathcal G_2(\mathbf{ k}_1,\mathbf{ k}_2,i\omega_q)=(1-f_{\mathbf {k}_1})/(i\omega_q-\varepsilon_{\mathbf k_1}-\varepsilon_{F\mathbf k_2}+\mu_F),$ with $f_{\mathbf {k}}=[\exp\beta(\varepsilon_{F\mathbf k_2}-\mu_F)+1]^{-1}$ the Fermi distribution. This  accounts for the 
Fermi blocking of the intermediate scattering states. Despite its simplicity, the ladder approximation turns out to be surprisingly accurate for neutral impurities 
in a Fermi gas~\cite{Massignan2014}. This can be attributed to an almost perfect cancellation of higher order particle-hole excitations  due to  Fermi statistics~\cite{Combescot2008}.
While such a cancellation may be less complete for the longer range atom-ion potential,
 we expect the ladder approximation to describe correctly the most important properties of mobile ions in a Fermi gas.

 \begin{figure}[!htb]
\centering
    \includegraphics[width=\columnwidth, height=0.8\columnwidth]{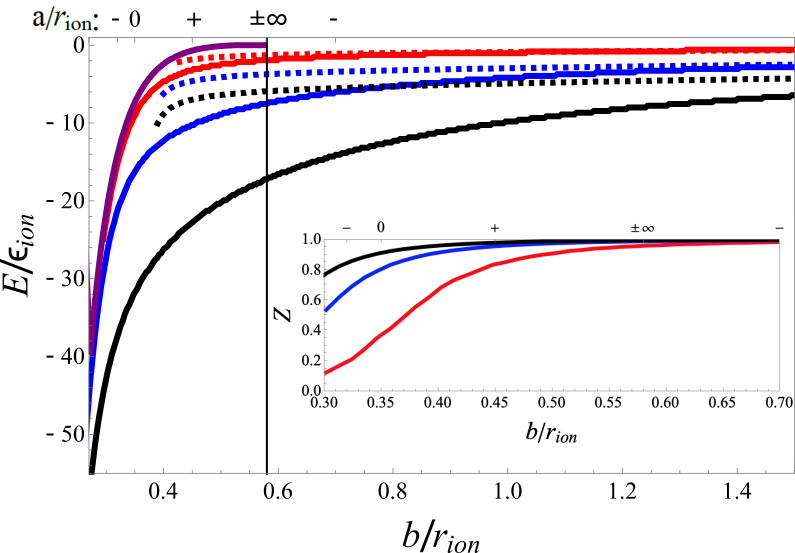}
    \caption{ Ionic polaron energy for the densities $n_Fr_\text{ion}^3=0.1$ (red),  $n_Fr_\text{ion}^3=0.5$ (blue), and  $n_Fr_\text{ion}^3=1.0$ (black). The dashed 
    lines give the energy of the neutral polaron for the same scattering length and densities, and 
    the purple line gives the binding energy of the highest molecular ion. 
    (Inset) Residue of the ionic polaron for the same densities as in the main panel using the same 
    color scheme.}
    \label{Fig_polvsmf}
\end{figure}
We start by studying the regime where there is at most one bound state in the atom-ion potential. 
In Fig.~\ref{Fig_polvsmf}, we show the  zero momentum quasiparticle energy $\varepsilon_P$ determined by the pole of the ion Green's function Eq.~\eqref{Gion}  taking 
$i\omega\rightarrow \omega+i0_+$, i.e.\ solving $\varepsilon_P=\Sigma(\mathbf 0,\varepsilon_P)$. 
The energy is plotted for three different densities $n_Fr_\text{ion}^3=0.1, 0.5$, and $1.0$ where $n_F=(2m_F\mu_F)^{3/2}/6\pi^2$, 
as a function of $b$ with the corresponding scattering length shown on top. 
This quasiparticle is the charged analogue of the  Fermi polaron observed for neutral impurities in atomic gases~\cite{Schirotzek2009,Kohstall12,Koschorreck12,Scazza2017,Fritsche2021},
and we  see that its energy decreases with increasing depth of the 
atom-ion potential and with the Fermi density $n_F$ as expected.  The purple line 
gives the energy of the first bound state, which emerges for $b/r_{\text{ion}}\simeq 0.58$ indicated by a vertical black line. 
For comparison, we also show the  energy obtained using the ladder approximation for a short-range interaction with the same scattering length $a$. 
Physically, this corresponds to the energy of the Fermi polaron for a neutral impurity where the interaction is short ranged and well described by the scattering length only~\cite{Massignan2014}.
We stop plotting the energy of this neutral polaron when its residue is below $0.05$. 
 For low density $n_Fr_\text{ion}^3=0.1$, the energy of the charged polaron is close to that of the neutral polaron. This reflects that the average distance between the particles is much larger than the characteristic length $r_\text{ion}$ of the atom-ion interaction so that it is well described by the scattering length only. The agreement however breaks down for strong interaction 
 close to $b/r_\text{ion}\simeq 0.58$  where a bound state emerges and  $1/a=0$. Here, the atom-ion correlations are strong 
 and depend on the range of the interaction. For larger density, the energy of the ionic polaron is 
 considerably smaller than that of the neutral polaron  except in the weakly interacting  
 regime $n_F^{1/3}a\ll 1$, where it is given by the mean-field value $2\pi an_F/m_r$.
 This demonstrates that the scattering length approximation is insufficient when the average interparticle spacing is comparable to or smaller than the range the atom-ion interaction.  It is also consistent
  with what is found for a mobile ion in a BEC~\cite{Christensen2021,Astrakharchik:2021wl}.
 
In the inset of Fig.~\ref{Fig_polvsmf}, we show the quasiparticle residue  obtained as 
$Z_{\mathbf p}^{-1}=1-\partial_\omega \text{Re}\Sigma(\mathbf p,\omega)|_{\varepsilon_P}$ for the same values of densities as  in the main figure. Surprisingly, we see that the residue of the polaron 
\emph{increases} with density. Intuitively, one would expect that the residue decreases since interaction effects should increase with density of the surrounding bath. 
Indeed, one finds that the residue of a  neutral polaron interacting with a short-range potential decreases with increasing density. 
We also see from  Fig.~\ref{Fig_polvsmf} that 
the polaron is a well-defined quasiparticle with a non-zero residue even for $a\rightarrow 0_+$,  in stark contrast to the case of a neutral
attractive polaron, where the residue vanishes in this so-called BEC limit~\cite{Massignan2014}. These two effects can  be understood from the fact that  the energy  
$\varepsilon_M$  of the molecule present 
for $b/r_\text{ion}< 0.58$ is higher than the value $-1/2m_ra^2$ valid for a short range potential. In particular, it is finite for $a\rightarrow0_+$ as can be seen from Fig.~\ref{Fig_polvsmf}. It follows that for a large density $\epsilon_F/\varepsilon_M$ remains significant even when $a\rightarrow0_+$ leading 
to a non-zero residue of the ionic polaron. For a small density on the other hand, $\epsilon_F/\varepsilon_M\ll1$  for $a\rightarrow0_+$ similar to a short range interaction 
leading to a small polaron residue. 
Equivalently, for a small density the polaron energy quickly approaches that of the molecule leading to a small residue, whereas the energy separation between the polaron 
and the molecule increases with the density leading to a larger residue. 
 Fig.~\ref{Fig_polvsmf} therefore showcases that the long-range aspects of the ion-atom potential  alter the properties of the ionic polaron compared to the neutral polaron.

\begin{figure}[!htb]
\centering
    \includegraphics[width=\columnwidth,height=0.85\columnwidth]{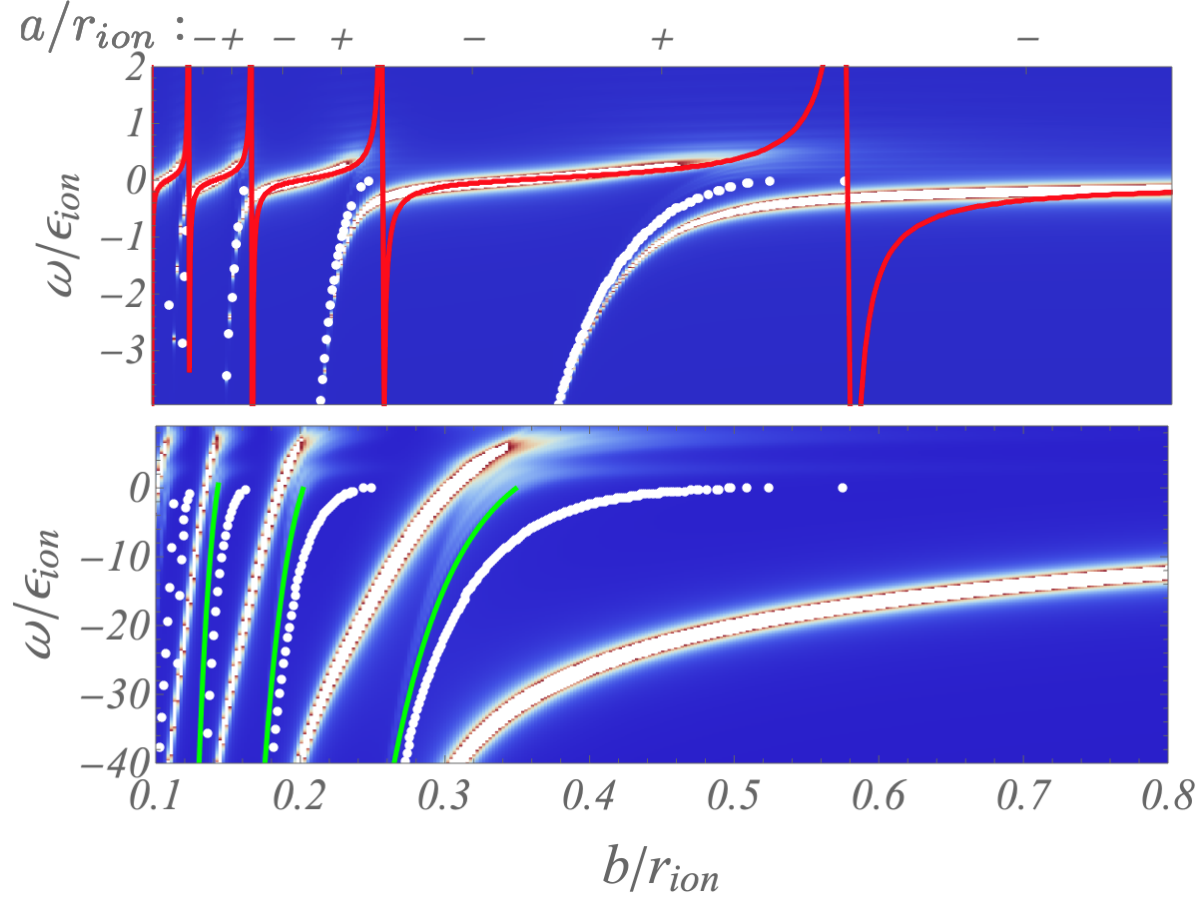}
    \caption{Zero momentum spectral function of an ion. (top) Low-density regime corresponding to $n_Fr_\text{ion}^3=0.01,$ and high-density regime $n_Fr_\text{ion}^3=1.0$ in (bottom). The white dots depict the binding energy $\epsilon_B^{(0)}$, the red line corresponds to the mean-field energy. The molecule energy in presence of the Fermi sea is shown by the green lines (bottom).}
    \label{Fig_hdvsld}
\end{figure}

To explore further the interplay between two- and many-body physics, we show in Fig.~\ref{Fig_hdvsld} the zero momentum
ion spectral function defined as 
$A(\mathbf p,\omega)=-2\text{Im}\mathcal G(\mathbf p,\omega)$ as a function of $b$  both for 
 low- and high-density with $n_Fr_\text{ion}^3=0.01$ and $n_Fr_\text{ion}^3=1$. We see that each time the atom-ion interaction potential 
 supports another bound state, a new polaron branch emerges. It follows  that when there are 
 $N$ molecular states, there are $N+1$ polaron branches in analogy with what has been 
 predicted for a mobile ion in a BEC~\cite{Massignan2005,Christensen2021}. However, the spectral weight of the low energy branches quickly 
 decrease as the binding energy of the molecular states increase. This is especially apparent in the low density regime  $n_Fr_\text{ion}^3=0.01$ where the 
 Fermi energy quickly becomes much smaller than the molecular binding energies corresponding to the BEC limit where the attractive polarons lose their spectral weight 
 consistent with what we discussed above in connection with Fig.~\ref{Fig_polvsmf}. The spectral weight is dominated by the highest energy polaron, which has 
 an energy that agrees  well with the mean-field prediction $2\pi n_Fa/m_r$ (red line) whenever $n_F|a|^3\ll 1$. This also 
 means that it  smoothly evolves from a repulsive $\varepsilon_P>0$ to an attractive $\varepsilon_P<0$ polaron with 
 decreasing $b$. By comparing the two panels in Fig.~\ref{Fig_hdvsld}, we see that the emergence of the new polaron branch is pushed to smaller values of $b$ for 
 high density  $n_Fr_\text{ion}^3=1.0$ and that the lower polaron branch  retains its spectral weight correspondingly longer. 
 Also, the  energy of the highest polaron quickly deviates from the mean-field prediction even when  $n_F|a|^3\ll 1$, again illustrating the breakdown 
 of the scattering length description when the interparticle distance becomes comparable to $r_\text{ion}$.
Note that  the effective interaction potential given by Eq.~\eqref{potential}   can only be expected to reproduce the correct
energy of the highest bound  $s$-wave state of the real atom-ion interaction. Thus, while  the general behaviour shown in Fig.~\ref{Fig_hdvsld} is robust, the spacing between the molecular resonances will depend on the specific atom-ion combination.

\begin{figure}[!htb]
\centering
    \includegraphics[width=\columnwidth]{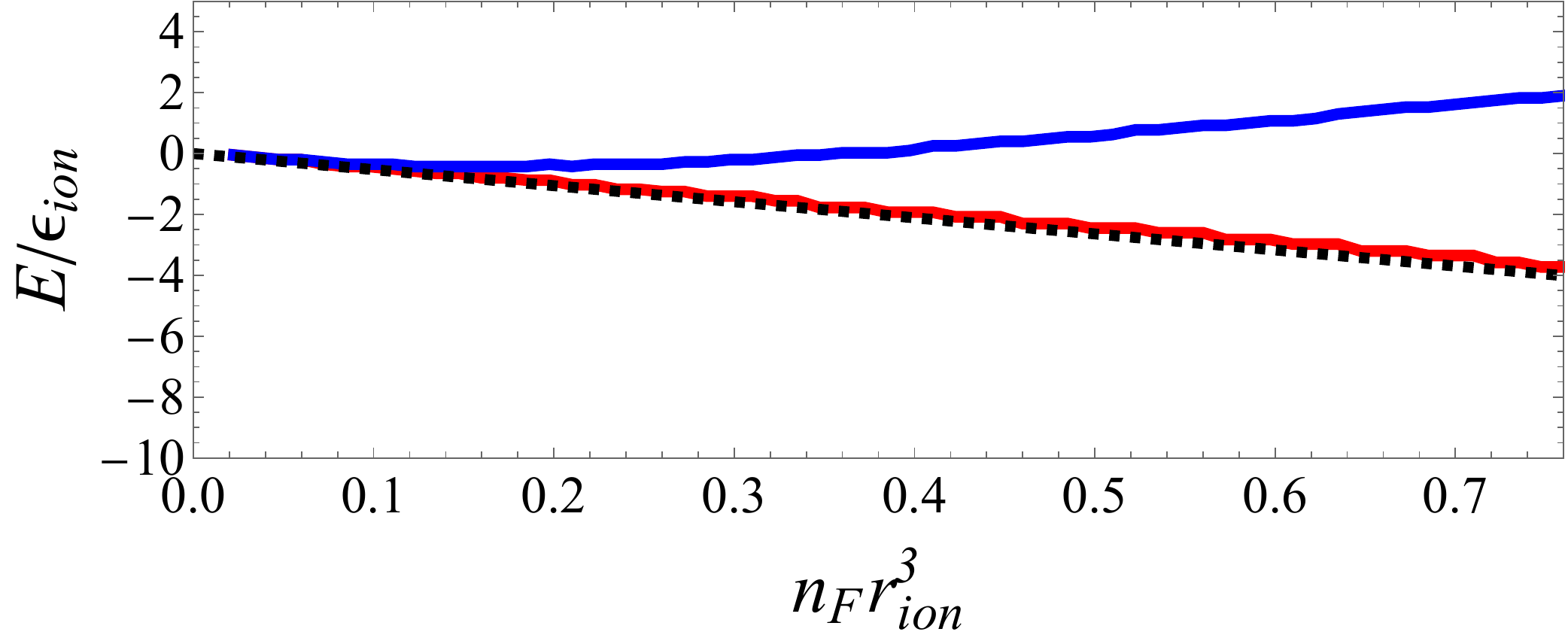}
    \caption{The energy of the highest polaron branch as a function of density for  $b/r_\text{ion}=2.0$ (red) and $b/r_\text{ion}=0.32$ (blue), which both 
    correspond to the scattering length $a/r_\text{ion}=-0.42$. The mean-field prediction is illustrated by the dashed black line.}
    \label{FigC}
\end{figure}

 We plot in Fig.~\ref{FigC} the energy of the zero momentum  polaron as a function of the density 
 for 	$b/r_\text{ion}=2$ and $b/r_\text{ion}=0.3237$ (the highest polaron branch), which both correspond to the  scattering length  $a/r_\text{ion}\simeq -0.42$. 
 For low densities, the two polaron energies are accurately described  by the mean-field expression $\varepsilon_P=2\pi a n_F/m_r$ (black dashed line). 
 The agreement breaks down with increasing density for the polaron  at $b/r_\text{ion}=0.3237$. 
Such different properties of the two highest polaron branches sharing the same scattering length can be explained as follows. There is only one polaron branch at $b/r_\text{ion}=2$ and 
no molecular state, whereas the highest polaron for $b/r_\text{ion}=0.3237$ is above both a molecular state and a lower polaron branch. As the density increases, this
upper polaron branch is therefore pushed away from the molecular state and the lower polaron branch. Since these effects are not included in mean-field theory, it has a smaller regime of validity for the 
highest polaron branch at   $b/r_\text{ion}=0.3237$ compared to that at  $b/r_\text{ion}=2$.

\section{Ionic molecule}
\label{molecule}
We finally turn our attention to  medium effects on the molecular ions.  Inspired by a similar approach for neutral impurities in a Fermi gas~\cite{Mora2009,Punk2009,Combescot2009,Massignan2014}, 
we can write down  the variational ansatz 
\begin{align}
\label{dimeransatz}
    |\psi_M\rangle = \left(\sum_{\mathbf{k}} \phi_\mathbf{k} \hat{c}^\dagger_{\mathbf{k}} \hat{a}^\dagger_{\mathbf{-k}}
    + \sum_{\mathbf{kk'q}} \kappa_{\mathbf{kk'q}} \hat{c}_{\mathbf{k}}^\dagger \hat{c}^\dagger_{\mathbf{k'}} \hat{c}_{\mathbf{q}} \hat{a}^\dagger_{\mathbf{q-k-k'}}\right) |\text{FS}\rangle
\end{align}
to describe these  dimer states in the presence of the 
Fermi sea $|\text{FS}\rangle$ focusing on zero total momentum. Here, $\phi_\mathbf{k}$ and $\kappa_\mathbf{kk'q}$ are variational parameters, which are determined 
by minimising  the energy $\langle \psi_M|H|\psi_M\rangle$. The details of this calculation and the resulting variational equations for $\phi_\mathbf{k}$ and $\kappa_\mathbf{kk'q}$ 
 are presented  in the Appendix. For simplicity, we present results for the molecule energy including only the first term in Eq.~\eqref{dimeransatz} taking 
 $\kappa_{\mathbf{kk'q}}=0$.
 
 The energy of the molecular ion obtained from  Eq.~\eqref{dimeransatz} is plotted in the Fig.~\ref{Fig_hdvsld} as solid green lines. We see that the presence of the Fermi sea 
 increases the molecule energy as compared to the vacuum case, which can easily be understood in terms of Fermi blocking of the available states for the molecular wave function. 
 This effect therefore increases with the density, and we have  not plotted the  energy predicted by 
 Eq.~\eqref{dimeransatz} for the low density case $n_Fr_\text{ion}^3$ as it is indistinguishable from the molecule energy in a vacuum.

\section{Conclusions and outlook}
\label{Conclusion}
We studied a single mobile ion in a quantum degenerate single-component Fermi gas. By calculating the full spectral response of the ion, we showed that the  long-range atom-ion 
interaction gives rise to the presence of several ionic Fermi polaron branches, which have quite different properties as compared to neutral Fermi polarons. 
The ionic and neutral polarons were demonstrated to have similar properties  only in the dilute limit where the average interparticle spacing is large compared to the 
characteristic length of the atomic interaction. We   finally showed the effects of the Fermi gas on the atom-ion bound states states, i.e.\ the molecular ions. 

Our results illustrate the rich physics that can be realised in  hybrid ion-atom gases, and  motivate future work. 
In particular, it would be highly relevant to calculate the three-body recombination rate for molecule formation for an ion in a degenerate Bose and Fermi gas to explore the lifetime 
of these systems~\cite{PerezRios2015,Wang2017,Wang2019}. This could be compared to neutral Bose and Fermi gases, where the three-body rate has been shown to depend on the scattering length as a simple  power 
law~\cite{Fedichev1996,Nielsen1999,Esry1999,Bedaque2000,Weber2003,Petrov2003}. Also, while the ladder approximation has been shown to be quantitatively reliable for a neutral 
impurity in a Fermi gas, an interesting question  concerns its accuracy for the case of an ion where the long-range nature of the atom-ion interaction might give rise to 
significant corrections. Likewise, a remaining problem   concerns  the inclusion of more terms in the ansatz for the molecular ion, which might  lead to a crossing of its energy with the polaron energy,
 as is  the case for neutral impurities~\cite{Mora2009,Punk2009,Combescot2009}. Finally, it would be interesting to calculate the induced interaction between two ions mediated by density 
 oscillations in the surrounding medium, which  in the case of neutral impurities can be quite strong leading to the formation 
 of bound states~\cite{Camacho2018b,Camacho2018a}. An accurate way to measure such an induced interaction could be to measure its effect on the phonon modes of  ions confined in a
 radiofrequency or dipole trap~\cite{Feldker2020,Schmidt2020}.

{\it Acknowledgments.-} This work has been supported by the Danish National Research Foundation through the Center of Excellence “CCQ” (Grant agreement no.: DNRF156), the Independent Research Fund Denmark-Natural Sciences via Grant No. DFF -8021-00233B, and the U.S. Army CCDC Atlantic Basic and Applied Research via Grant No. W911NF-19-1-0403.

 \bibliography{references}

\newpage
\onecolumngrid
\appendix
\section*{Appendix}
\section{Dressed dimer equations}
We derive minimization equations for the dressed dimer with full ion potential. The Hamiltonian is still 
\begin{gather}
\hat{H} = \sum_{\mathbf{k}} \left(\varepsilon_{\mathbf k}^I\hat{f}^\dagger_\mathbf{k} \hat{f}_\mathbf{k} +\varepsilon_{F\mathbf k} \hat{c}^\dagger_\mathbf{k} \hat{c}_\mathbf{k}\right) + \sum_{\mathbf{k}, \mathbf{k'},\mathbf{q}} V(\mathbf{q}) \hat{c}^\dagger_{\mathbf{k'-q}} \hat{c}_{\mathbf{k'}} \hat{f}^\dagger_{\mathbf{k}+\mathbf{q}}
\hat{f}_\mathbf{k},
\end{gather}
while the ansatz can be conveniently written as
\begin{align}
    |\psi^{D} \rangle = \big( \underbrace{\sideset{}{'}\sum_\mathbf{k}\phi_\mathbf{k} \hat{c}^\dagger_\mathbf{k} \hat{f}^\dagger_{\mathbf{-k}}}_{\psi_0^D} + \underbrace{\sideset{}{'}\sum_{\mathbf{kk'q}} \kappa_{\mathbf{kk'q}} \hat{c}^\dagger_\mathbf{k} \hat{c}^\dagger_\mathbf{k'} \hat{c}_\mathbf{q} \hat{f}^\dagger_{\mathbf{q-k-k'}} }_{\psi_1^D} \big)|FS^N \rangle.
\end{align}
We calculate expectation values and show each term explicitly for the interested reader: 
\begin{align}
    \langle\psi_0^D | H_0 | \psi_0^D \rangle  & = \sideset{}{'}\sum_{\mathbf{k}} \phi_\mathbf{k}\phi_\mathbf{k}^\ast \left( F\epsilon_\mathbf{k} + \epsilon^I_\mathbf{k}-\epsilon_F \right)\\
    \langle\psi_1^D | H_0 | \psi_1^D \rangle  & = \sideset{}{'}\sum_{\mathbf{kk'q}} \left( \kappa_\mathbf{kk'q}\kappa_\mathbf{kk'q}^\ast -\kappa_\mathbf{kk'q}\kappa_\mathbf{k'kq}^\ast \right) \left( \epsilon^I_\mathbf{q-k-k'} + \epsilon_{F\mathbf{k}} + \epsilon_{F\mathbf{k'}} - \epsilon_{F\mathbf{q}}+ \sideset{}{'}\sum_\mathbf{q'}\epsilon_{F\mathbf{q'}} - \epsilon_F \right)\\
     \langle \psi_0^D | H_I | \psi_0^D \rangle & = \sideset{}{'}\sum_{\mathbf{kk'}} \phi_\mathbf{k} \phi_\mathbf{k'}^\ast V(\mathbf{k'-k}) + V(\mathbf{0})n_F \sideset{}{'}\sum_{\mathbf{k}}\phi_\mathbf{k}\phi_\mathbf{k}^\ast\\
     \langle \psi_1^D | H_I | \psi_1^D \rangle & = \sideset{}{'}\sum_\mathbf{k k' q q'} \left(\kappa_\mathbf{k k' q} \kappa_\mathbf{k k' q' }^\ast - \kappa_\mathbf{k k' q} \kappa_\mathbf{k' k q' }^\ast \right) V(\mathbf{q-q'}) + \sideset{}{'}\sum_\mathbf{k k_1 k_2 q} \left( \kappa_\mathbf{k_1 k q} \kappa_\mathbf{k_2 k q}^\ast-\kappa_\mathbf{k_1 k q} \kappa_\mathbf{k k_2 q}^\ast \right) V(\mathbf{k_2-k_1}) \nonumber \\
    &+ \sideset{}{'}\sum_\mathbf{k k_1 k_2 q} \left( \kappa_\mathbf{k k_1 q} \kappa_\mathbf{k k_2 q}^\ast-\kappa_\mathbf{k k_1 q} \kappa_\mathbf{k_2 kq}^\ast \right) V(\mathbf{k_2-k_1})+V(\mathbf{0})n_F\sideset{}{'}\sum_\mathbf{kk'q}\left(\kappa_\mathbf{kk'q} \kappa_\mathbf{kk'q}^\ast- \kappa_\mathbf{kk'q} \kappa_\mathbf{k'kq}^\ast\right)\\
     \langle \psi_1^D | H_I | \psi_0^D \rangle & = \sideset{}{'}\sum_\mathbf{kk'q} \kappa_\mathbf{kk'q}^\ast \left(\phi_\mathbf{k'} V(\mathbf{q-k}) - \phi_\mathbf{k} V(\mathbf{q-k'})\right)  . 
\end{align}
We minimize by setting $\partial \phi_\mathbf{k}^\ast \langle (H- E) \rangle=0 $ and $\partial \kappa_\mathbf{kk'q}^\ast \langle (H- E) \rangle=0$ and we get the following coupled equations: 
\begin{align}
    (-E+\varepsilon_{F\mathbf{k}} + \varepsilon_\mathbf{k}^I+V(\mathbf{0})n_F-\epsilon_F)\phi_\mathbf{k} + \sideset{}{'}\sum_\mathbf{k'} V(\mathbf{k'}-\mathbf{k}) \phi_\mathbf{k'} + \sideset{}{'}\sum_{\mathbf{k' q }} \left(\kappa_{\mathbf{k k' q}}-\kappa_{\mathbf{k'kq}}\right) V(\mathbf{q-k'})=0,
\end{align}
and 
\begin{multline}
    -\phi_\mathbf{k_1} V(\mathbf{q-k_2})+\phi_\mathbf{k_2}V(\mathbf{q-k_1}) + E_\mathbf{k_1 k_2 q} (\kappa_\mathbf{k_1 k_2 q}-\kappa_\mathbf{k_2 k_1 q}) + \sideset{}{'}\sum_{\mathbf{q'}} (\kappa_\mathbf{k_1 k_2 q'}-\kappa_\mathbf{k_2 k_1 q'}) V(\mathbf{q'-q})\\ + \sideset{}{'}\sum_\mathbf{k} (\kappa_\mathbf{k_1 k q} - \kappa_\mathbf{k k_1 q}) V(\mathbf{k_2 - k}) +  \sideset{}{'}\sum_\mathbf{k} (\kappa_\mathbf{k k_2 q}-\kappa_\mathbf{k_2kq}) V(\mathbf{k_1-k})=0,
\end{multline}
where $E_\mathbf{k_1 k_2 q}= -E+\varepsilon_{F\mathbf{k_1}}+\varepsilon_{F\mathbf{k_2}-}\varepsilon_{F\mathbf{q}}+\varepsilon_\mathbf{q-k_1-k_2}^I+V(\mathbf{0})n_F-\epsilon_F$. The inclusion of the $\epsilon_F$ is because we still choose to the measure the energy $E$ in terms of the $N$ particle energy of the fermi sea, even though there are only $N-1$ particles in the fermi sea for the case of the dressed dimer. 




\end{document}